\documentclass{aa}

\usepackage{graphicx}
\usepackage{float}
\usepackage{sidecap} 
\usepackage{txfonts}
\usepackage{color}
\usepackage{braket}

\usepackage{pifont}
\usepackage[switch]{lineno}

\usepackage{enumitem}
\usepackage{ulem}

\usepackage{color}

\usepackage{tikz}

\usepackage{subcaption}

\usepackage[colorlinks=true,allcolors=blue]{hyperref}

\begin{document}


\title{Collisional rate coefficients for OH-H$_2$ at high temperatures}
   \author{Z. van den Heuvel \inst{1}
         \and
         B. Tabone \inst{2}
         \and
         E. F. van Dishoeck
          \inst{3,4}
         \and
         G. C. Groenenboom \inst{1}
         \and
         A. van der Avoird \inst{1}
}
          \institute{
        Theoretical Chemistry, Institute for Molecules and Materials,
        Radboud University, Heyendaalseweg 135, 6525 AJ Nijmegen, The
        Netherlands \\
        \email{A.vanderAvoird@theochem.ru.nl}
              \and
      Université Paris-Saclay, CNRS, Institut d’Astrophysique Spatiale,
91405 Orsay, France
       \and
      Leiden Observatory, Leiden University, P.O. Box 9513,
            2300 RA Leiden, The Netherlands
              \and
      Max Planck Institut f\"ur Extraterrestrische Physik (MPE), Giessenbachstrasse 1, 85748 Garching, Germany
}

\date{Received  2025; accepted yyy}

 \abstract{OH is a cornerstone molecule in the chemistry of interstellar and circumstellar media and is ubiquitously detected in warm gas thanks to its infrared rotational lines. However, the excitation processes of OH remain poorly characterized.}
 {We provide a new set of collisional rate coefficients for OH with H$_2$, expanding the existing data to $j$ levels up to $j=15/2$ and temperatures up to 750~K.}
 {These rate coefficients are obtained from state-to-state collision cross sections calculated by means of well-converged close-coupling quantum scattering calculations for collisions of OH with para- and ortho-H$_2$ with energies up to 1700~cm$^{-1}$ ($\simeq 2450$\,K).}
 {We reproduce the rate coefficients computed by Kłos et al. (2017) and extend their results to higher temperatures and higher rotational levels of OH. The de-excitation rate coefficients are lower in collisions with para-H$_2$ ($j_{\rm H_2} = 0$) due to the absence of a quadrupole moment, but this difference decreases at higher temperatures. We find that the rate coefficients follow scaling relations with the energy gap between the upper and lower levels of a given transition, which allows extrapolation to higher OH rotational states $j_{\rm OH}$. As a first application, we show that under astrophysical conditions typical of warm and dense gas around nascent stars, the populations of low-$j_{\rm OH}$ states are dominated by collisions, even when chemical pumping is included.
 The full set of rate coefficients is made available in the LAMDA database.}
 {OH infrared emission provides a unique probe of local conditions in astrophysical environments. These rate coefficients contribute to developing a complete excitation model of OH under warm conditions, with chemical pumping of OH through the O + H$_2$ reaction now being the main remaining uncertainty in such models.}
   \keywords{atomic and molecular data -- astrochemistry
               }

   \maketitle

%

\section{Introduction}
\label{sec:intro}
The hydroxyl radical, OH, is a central molecule in a wide variety of
interstellar environments, as shown by models and observations ranging
from shocked gas associated with star-forming regions and supernova
remnants
\citep[e.g.,][]{Hollenbach89,Hollenbach89ESA,Draine83,Reach98,Godard19}
to massive outflows in extragalactic systems
\citep{Sturm11,Gonzalez12}, to warm gas in the envelopes of evolved
stars \citep[e.g.,][]{Nejad88} and in the inner regions of
planet-forming disks around nearby stars
\citep[e.g.,][]{Carr08,Salyk08,Glassgold09}. Its importance is twofold.
First, it informs on the partitioning between the major forms of
oxygen in warm interstellar gas: O, OH and H$_2$O
\citep[e.g.,][]{Neufeld89,Walsh15}.  The three species are linked
through the O + H$_2$ $\to$ OH + H and OH + H$_2$ $\to$ H$_2$O + H
reactions which proceed rapidly once the temperature is above
$\simeq$300 K \citep{Charnley97}. Backward reactions with H and UV
photodissociation drive the chemistry in the other direction. Second,
OH and H$_2$O are important coolants of the gas because of their rich
far-infrared line spectra \citep[e.g.,][]{Hollenbach79,Kaufman96}.  In
order to quantitatively assess the role of OH in the chemistry and
physics in interstellar gas, its excitation needs to be understood in
order to translate the observed line intensities to column densities,
temperatures and densities. Collisional rate coefficients are a key
ingredient to compute its excitation.

To properly describe previous work on OH in the remainder of this introduction, the quantum labels of the OH states first need to be summarized. The exact quantum numbers labeling the levels of OH in its $^2\Pi$
electronic ground state are the total angular momentum $j$, the parity
$p$ [or the spectroscopic parity $\epsilon = p \,(-1)^{j-1/2}$], and the
labels $F_1$ and $F_2$ of the lower and upper spin-orbit states. The
levels can also be characterized with different approximate quantum
numbers, either by the rotational and electronic angular momenta $N = j
\pm 1/2$ and $\Lambda = \pm 1$ in a Hund's case (b) description, or by
the spin-orbit quantum number $\Omega = 3/2$ or 1/2 in a Hund's case (a)
approach \citep{herzberg:50}. The case (a) approximate quantum number
$\Omega$ is best for the lower $j$ states of OH, the case (b) quantum
number $N$ is better for higher $j$. Analysis of the OH states shows,
however, that $N$ is a fairly good quantum number even for the lowest
state of mixed character with $j=3/2$: the eigenstates in the lower
spin-orbit manifold $F_1$ have 88\% of $N=1$ character, while the upper
spin-orbit states $F_2$ have 88\% of $N=2$ character. Each OH level with
given $N$ is split by spin-orbit coupling into two ladders with $\Omega
= 3/2$ and 1/2 named $F_1$ and $F_2$, respectively, which are further
split by the much smaller $\Lambda$-doubling into two sets of states
labeled by the spectroscopic parity $e/f$ for $\epsilon = +/-$1. An
alternative notation \citep{alexander:88} for parity is the level's $A'$
and $A''$ symmetry (with respect to reflection about the plane of
rotation of the molecule) to designate states with $\Omega$ = 1/2, $f$
or $\Omega$ = 3/2, $e$ and $\Omega$ = 1/2, $e$ or $\Omega$ = 3/2, $f$,
respectively. Thus, each $N$ $\to$ $N-1$ transition consists in
principle of four lines. These pure rotational intra- and cross-ladder
transitions occur at far-infrared wavelengths (see Figure 1 in \citealt{Tabone21} for energy level structure). Hyperfine structure due
to the spin $I=1/2$ of the H nucleus further splits each
$\Lambda$-doublet level into two levels.

OH was first detected in the interstellar medium at radio wavelengths
through its $\Lambda$-doubling transitions at 18 cm
\citep{Weinreb63}. These bright radio lines have since been widely
used to trace various astrophysical objects throughout our and other
galaxies \citep[e.g.,][]{Lo05,Engels15,Beuther19,Jarvis24}, However,
since these lines are amplified due to masering, their ability to
probe interstellar chemistry is limited.

Following its radio detection, OH has indeed been seen through thermal
emission or absorption at other wavelengths. In particular, its
lower-lying pure rotational transitions at far-infrared wavelengths
were detected 40 years ago in the bright Orion shocked region with
pioneering airborne instruments
\citep[e.g.,][]{Storey81,Watson85,Melnick87}.  In the following
decades, the Infrared Space Observatory (ISO)
\citep[e.g.,][]{Giannini01,Nisini02} and most notably the {\it
  Herschel} Space Observatory provided much more sensitive, high
spectral and spatial resolution data, both in galactic star-forming
regions
\citep[e.g.,][]{Wampfler10,Wampfler11,vanKempen10,Goicoechea15,Karska18},
photodissociation regions \citep[e.g.,][]{Goicoechea11,Parikka17},
protoplanetary disks \citep[e.g.,][]{Fedele13} and extragalactic
sources \citep[e.g.,][]{Gonzalez17}. These observations typically
cover OH lines arising in the $^2\Pi_{3/2}$ and $^2\Pi_{1/2}$ ladders with
upper energies $E_{\rm up}/k_B$ up to 1000 K, or $N\simeq$5.

The {\it Spitzer} Space Telescope, with its infrared spectrometer
covering 10--38 $\mu$m at $R=\lambda/\Delta \lambda=600$, surprisingly
detected much more highly excited OH emission in a few sources
originating from energy levels up to $N$=34 ($E_{\rm up}/k_B$=28\,000 K)
\citep{Tappe08,Carr14}. At such high energies other processes than
thermal excitation must play a role in populating the levels.  The
{\it James Webb} Space Telescope (JWST) with its superb sensitivity
and improved spectral resolving power ($R\simeq 3000$) at $1-28$
$\mu$m has opened up a new chapter in studies of OH, being able to
probe the stronger intra-ladder OH transitions with $N\geq 10$. JWST
now routinely detects many highly excited mid-infrared OH lines in
protoplanetary disks
\citep[e.g.,][]{Gasman23,Schwarz24,Arulanantham25} and dense shocks
associated with outflows
\citep[e.g.,][]{Francis24,Caratti24,vanGelder24}.  Indeed, OH lines up
to $N$=45 ($E_{\rm up}/k_B$=46\,500 K) have been detected with the
Mid InfraRed Instrument (MIRI) \citep{Wright23}, showing a very
characteristic pattern at 9--10 $\mu$m where the high-$N$ levels pile
up \citep[e.g.,][]{Zannese24,Neufeld24,Vlasblom25}.  Moreover, the
higher spectral resolution of MIRI allows the quartet of lines to
be (partially or fully) resolved depending on $N$ value, something
that {\it Spitzer} could not do.
Ro-vibrational lines of OH at 3 $\mu$m have also been
detected, both with JWST \citep{Zannese24} and using ground-based
telescopes \citep{Salyk08,Fedele11}. OH is clearly located in warm gas that
has temperatures ranging from a few hundred up to $\simeq$1500 K.

These OH detections have triggered renewed investigations into the
mechanisms that excite OH. The lower-lying rotational levels are
populated primarily by thermal collisions, which has stimulated various
theoretical studies into the rate coefficients of OH with its main collision
partner, H$_2$ \citep[e.g.,][]{Offer94,Klos20}. However, the OH
($v,J$) levels can also be populated by two other ``chemical pumping''
processes. First, the chemical reaction of O + H$_2$ produces OH with
a distribution over its $v, J$ levels \citep{Balakrishnan04}. However,
state-to-state reaction rates (i.e., considering the state
distribution of both the reagents and the products) have not yet been
published \citep{Veselinova2021}. Based on current estimates, this reaction is expected to be
most efficient in populating the mid-$J$ levels shortward of 20 $\mu$m
($N\simeq 14-20$).

Second, the photodissociation of H$_2$O produces OH with excess energy
that can be distributed over the $v, J$ states. This so-called
``prompt emission'' depends on the electronic state that is excited
\citep{Carr14,Tabone21}: H$_2$O photodissociation through the
$\tilde A$ excited electronic state ($\lambda$=143--200 nm) produces
vibrationally hot but rotationally cold molecules
\citep[e.g.,][]{Hwang99,vanHarrevelt01}. By contrast,
photodissociation through the $\tilde B$ state ($\lambda$=114--143 nm,
including Lyman $\alpha$ at 121.6 nm) produces OH in very high
rotational states up to $N\simeq 47$ \citep{Harich00,vanHarrevelt00},
just as is now being observed with JWST.  The origin of the very high
$N$ levels through H$_2$O photodissociation is confirmed by the fact
that this process exclusively produces OH in rotational states with an
$A'$ symmetry \citep{Zhou15}, i.e., $\Omega$=1/2, $f$ and $\Omega$=
3/2, $e$ labels. With the high spectral resolution of JWST, this
``smoking gun'' asymmetry in the doublet is now clearly seen, with
$A'/A''$ ratios $\geq 10$ \citep[]{Zannese24,Neufeld24}. However, a small
contribution from collisions would show up in emission from the $A''$
levels. Thus, a good understanding of the OH excitation can in this case
help to better constrain both the water abundance as well as the UV
environment and density of the gas in which OH is located.

The new JWST data have stimulated the development of more complete OH
excitation models that include both collisional excitation as well as
the chemical processes \citep[e.g.,][]{Tabone24,Neufeld24}. There is
now a clear need to include collisional rate coefficients of OH with
H$_2$ up to higher temperatures and $N$ values in the models, since
the mid-$N$ levels seen by JWST can be populated by a combination of
collisions and chemical pumping in high density and temperature gas
such as present in shocks and protoplanetary disks. Establishing up to
what $N$ levels collisions contribute to the excitation will be
important to infer the physical parameters of the gas in which OH
resides as well as its column density and abundance.

There have been a number of fundamental chemical-physics studies of
OH-H$_2$ collisions, dating back to the seminal experiments by
\citet{Andresen84} and \citet{Schreel96}. Stimulated by these
experiments and new observational data from ISO, \citet{Offer92}
computed collisional cross sections with both para- ($j=0$) and
ortho-H$_2$ ($j=1$) as collision partners but went up to only
$\simeq$300 K with $N$ up to 4. Hyperfine structure for low $N$ was
included by \citet{Offer94} and by \citet{Klos20} using new potential energy
surfaces. \citet{Schewe15} subsequently showed good agreement between
new theoretical cross sections for the lowest $\Omega=3/2$ transitions
and experimental data from \citet{Kirste10} over a 100--500 cm$^{-1}$
collision energy range. This work demonstrates the high level of
accuracy that can now be attained with theoretical methods.

In this work, we use quantum scattering methods to provide OH-H$_2$
collisional cross sections and rate coefficients up to higher energies
and $N$ levels, by extending the calculations of \citet{Schewe15}.
We then use an excitation model including collisions with H$_2$ and
chemical pumping to illustrate the importance of collisions in the
analysis of the OH pure rotational lines. This paper is organized as follows.
Section \ref{sec:cross} describes the calculation of the cross sections
and rate coefficients, the results, and an extrapolation procedure to
extend the results for $j_{\rm OH} \le 15/2$ to higher rotational
levels, Section \ref{sec:appl} presents an illustrative application, and Section
\ref{sec:concl} the Conclusions. The resulting rate coefficients are
made publicly available through the LAMDA database
\citep{Schoeier05,vanderTak20}.

\section{State-to-state cross sections and rate coefficients}
\label{sec:cross}
\subsection{Computational procedure}
\label{sec:comp}
We calculated de-excitation cross sections for OH-H$_2$ collisions with
OH in initial states with rotational angular momentum $j$ ranging from
1/2 to 15/2 and H$_2$ with angular momentum $j=0$ for para-H$_2$
(pH$_2$) and $j=1$ for ortho-H$_2$ (oH$_2$). Higher states of H$_2$
are also populated at the temperatures we consider, but we found after tests
with initial $j=2$ for pH$_2$ and $j=3$ for oH$_2$ that the
rate coefficients for higher initial $j$ of H$_2$ can be reasonably
accurately estimated from those calculated for $j=1$, hence they are not computed explicitly (see also below). For each initial
$j$ of OH we start from both the spin-orbit states $F_1$ and $F_2$ with
$\Omega=3/2$ and 1/2, respectively, and both parities $p=\pm 1$. The
parity $p$ is related with the spectroscopic parity $\epsilon$ as $p =
(-1)^{j-S}\epsilon$ with $S=1/2$ being the spin of the OH ground state.
States with $\epsilon=+1$ and $-1$ are labeled $e$ and $f$,
respectively. Hyperfine structure is not included.

For each initial state $\ket{j_{\rm OH},\Omega,\epsilon, j_{{\rm H}_2}}$
we compute the de-excitation cross sections for collisional transitions
to all final states $\ket{j'_{\rm OH},\Omega', \epsilon', j'_{{\rm
H}_2}}$ that are lower in energy. We used the numerically exact
coupled-channels or close-coupling method with renormalized Numerov
propagation \citep{johnson:78,johnson:79} to solve the coupled-channels
equations in body-fixed coordinates on a grid of OH-H$_2$ distances $R$
ranging from 3.2 to $30~a_0$ with 307 equidistant points ($a_0$ is the
Bohr radius of 0.529 \AA). At the largest $R$-value we then transformed
the renormalized Numerov wave function quotient matrix $Q$ from
body-fixed to space-fixed coordinates and applied the appropriate
boundary conditions to obtain the scattering matrices $S$ from which the
cross sections are calculated. This procedure, which is computationally
efficient, is implemented in the Nijmegen coupled-channels program
described by \citet{jongh:17}, who applied it to NO-H$_2$. We calculated
the cross sections for collision energies ranging from 1 to
1700~cm$^{-1}$. At lower energies we find peaks in the cross sections
due to scattering resonances. Therefore, we used a logarithmic energy
grid with 50 points in the range from 1 to 500~cm$^{-1}$ and an equally
spaced grid with steps of 100~cm$^{-1}$ from 500 to 1700~cm$^{-1}$.

We carefully checked convergence of the cross sections with respect to
the $R$-grid, the channel basis of OH functions with maximum $j=25/2$
and H$_2$ functions with maximum $j=5$, and the total angular momentum
of the collision complex with maximum $J=159/2$. These maximum values
were needed for the highest initial OH state $j=15/2(F_2)$ and the
highest collision energy, for lower initial states and lower collision
energies we could use smaller values and a smaller $R$-grid. The
rotational constants of OH and H$_2$ were taken as 18.5487 and
59.3398~cm$^{-1}$, respectively. The spin-orbit coupling constant of OH
is $-139.21$~cm$^{-1}$ and the $\Lambda$-doubling parameters are $p =
0.235$~cm$^{-1}$ and $q = -0.0391$~cm$^{-1}$.

We used the set of diabatic OH-H$_2$ potentials of \citet{ma:14} from
their RCCSD(T)-F12a calculations. They truncated the potential to a
constant value for $R < 3.5~a_0$. For higher collision energies we
needed to start the propagation at smaller $R$ to obtain convergence, so
we linearly extrapolated the repulsive wall at small $R$ down to
$R=3.0~a_0$. At a collision energy of 2000~cm$^{-1}$ the differences in
the cross sections calculated with and without the small-$R$
extrapolation of the potential were smaller than 0.1\%, however. The
work by Schewe et al.\ \citep{Schewe15} has shown that calculated cross
sections in OH collisions with para- and ortho-H$_2$ for energies in the
range from 75 to 150~cm$^{-1}$ agree with crossed molecular beam results
to within the experimental error bars. Since their calculations were
also based on the OH-H$_2$ potential of Ma et al.\ \citep{ma:14} it
shows that this potential is accurate.

Temperature-dependent state-to-state rate coefficients were obtained for
$T=10$ to 750\,K from the energy-dependent cross sections
$\sigma_{j'_{\rm OH}\Omega' \epsilon' j'_{{\rm H}_2} \leftarrow
j_{\rm OH}\Omega \epsilon j_{{\rm H}_2}}(E)$ with the formula
\begin{eqnarray}
\label{eq:rate}
 && k_{j'_{\rm OH}\Omega' \epsilon' j'_{{\rm H}_2}
 \leftarrow j_{\rm OH}\Omega \epsilon j_{{\rm H}_2}}(T) =
  \left(\frac{8 k_{B} T}{\pi \mu}\right)^{1/2}
\\
  &&\times\int_{0}^{\infty}
 \sigma_{j'_{\rm OH}\Omega' \epsilon' j'_{{\rm H}_2} \leftarrow
    j_{\rm OH}\Omega \epsilon j_{{\rm H}_2}}(E) \left(\frac{E}{k_B T}\right) \,
\exp\left(-\frac{E}{k_B T}\right)\, d\left(\frac{E}{k_B T}\right),
\nonumber
\end{eqnarray}
where $k_B$ is the Boltzmann constant and $\mu$ is the reduced mass of
OH-H$_2$. The integration over energies was carried out numerically with
the Simpson rule, after interpolation of the cross sections on the
energy grid. The collision energy of 1700~cm$^{-1}$ (2450\,K) is not
sufficiently high to obtain the rate coefficients for temperatures up to
750\,K. We found, however, that even at 750\,K the integrand in
Eq.~(\ref{eq:rate}) already decays exponentially at the highest
energies, so we extrapolated the integrand with exponential functions
$a\exp(-b E)$ with constants $a$ and $b$ obtained from the largest two
energies and integrated over energies up to 5000~cm$^{-1} (7200\,K)$.

State-to-state excitation rate coefficients can be obtained from the
calculated de-excitation rate coefficients with the aid of the detailed
balance condition
\begin{eqnarray}
\label{eq:balance}
&& k_{j_{\rm OH}\Omega \epsilon j_{{\rm H}_2} \leftarrow j'_{\rm OH}\Omega' \epsilon' j'_{{\rm H}_2} }(T)
= \frac{(2 j_{\rm OH} + 1)(2 j_{{\rm H}_2} + 1)}{(2 j'_{\rm OH} + 1)(2 j'_{{\rm H}_2} + 1)}
\\
&&\times\exp\left(\frac{E_{j'_{\rm OH}\Omega' \epsilon' j'_{{\rm
H}_2}}-E_{j_{\rm OH}\Omega \epsilon j_{{\rm H}_2}}}{k_B T}\right)
k_{j'_{\rm OH}\Omega' \epsilon' j'_{{\rm H}_2} \leftarrow j_{\rm OH}\Omega \epsilon j_{{\rm H}_2} }(T),
\nonumber
\end{eqnarray}
with $E_{j_{\rm OH}\Omega \epsilon j_{{\rm H}_2}} = E_{j_{\rm OH}\Omega \epsilon} + E_{j_{{\rm H}_2}}$.
We checked the accuracy of this procedure by comparison with some
explicitly calculated excitation rate coefficients.

\subsection{Results}
\label{sec:results}

\begin{figure*}
\begin{tabular}{cc}
\includegraphics*[width=0.48\textwidth]{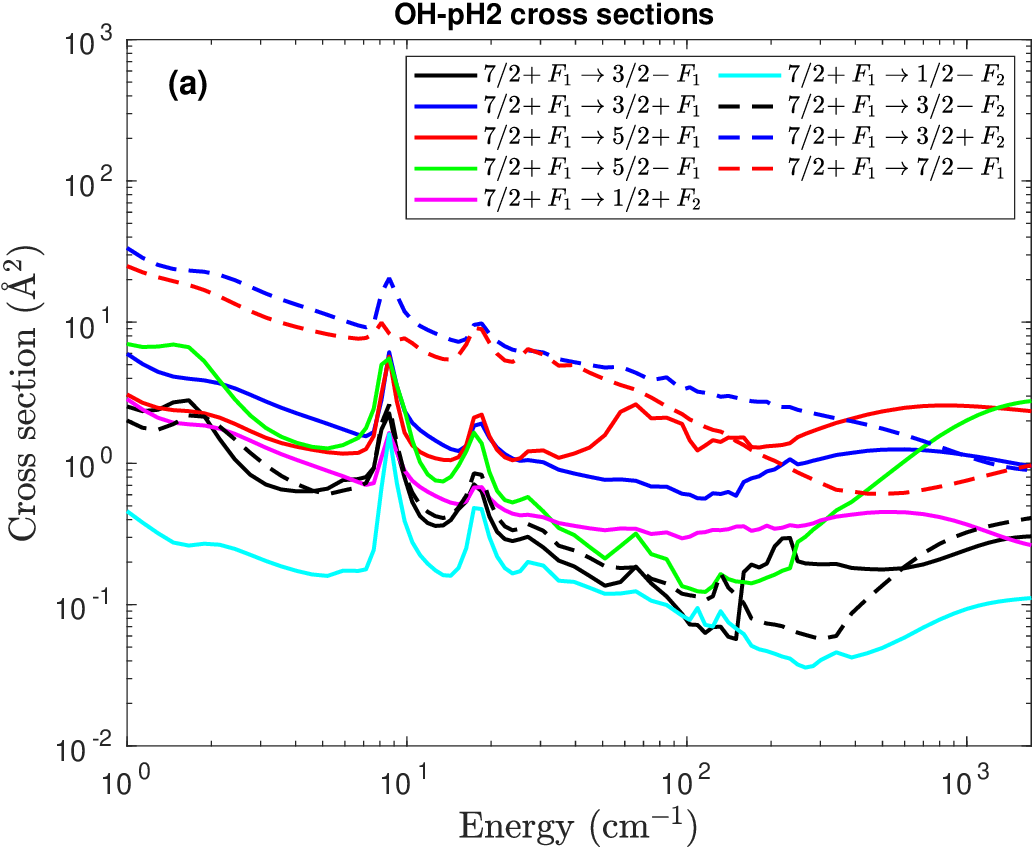} &
\includegraphics*[width=0.48\textwidth]{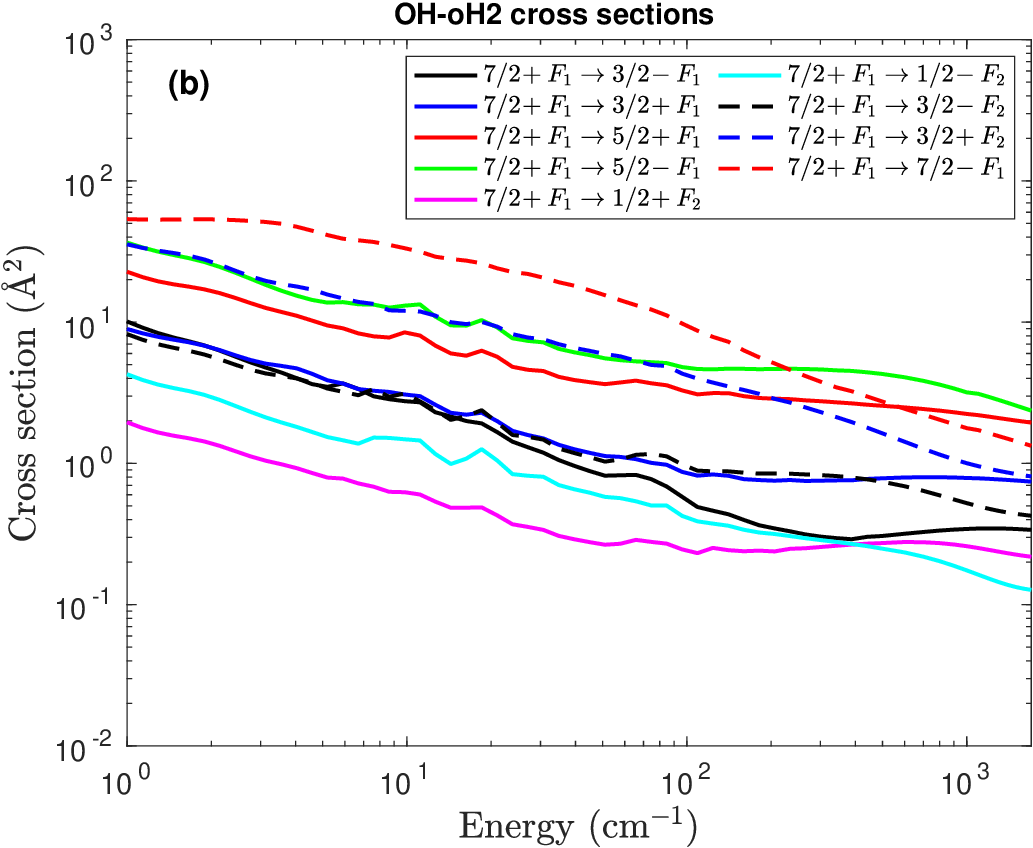} \\
\includegraphics*[width=0.48\textwidth]{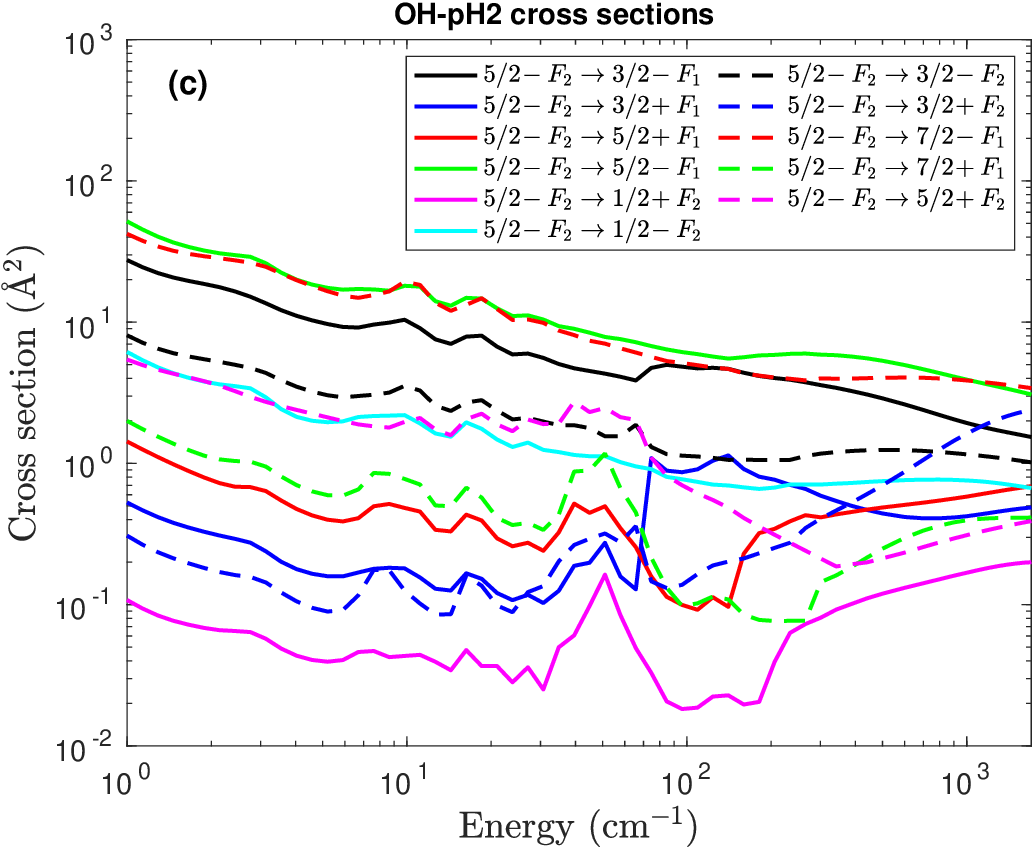} &
\includegraphics*[width=0.48\textwidth]{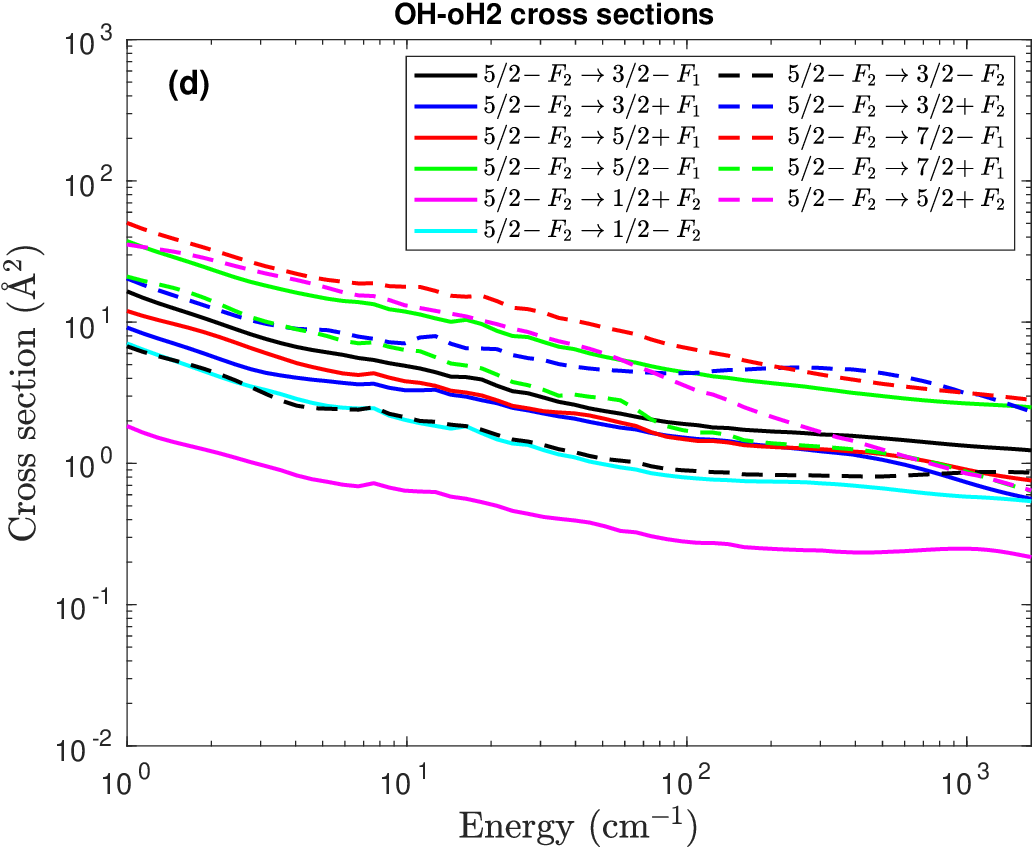} \\
\end{tabular}
\caption{Cross sections of OH transitions from the $F_1$ state with
$j=7/2$ and parity $p=+1$ (spectroscopic parity $f$) [panels (a) and
(b)] and from the $F_2$ state with $j=5/2$ and parity $p=-1$
(spectroscopic parity $f$) [panels (c) and (d)] to all lower OH levels.
H$_2$ is assumed to be initially in $j=0$ for pH$_2$ and in $j=1$ for
oH$_2$ and the cross sections are summed over all final states of
pH$_2$ and oH$_2$, respectively. Panels (a) and (c) refer to collisions
with pH$_2$, panels (b) and (d) to collisions with oH$_2$. The final
states of OH are listed in order of increasing energy.}
\label{fig:cross}
\end{figure*}

\begin{figure*}
\begin{tabular}{cc}
\includegraphics*[width=0.48\textwidth]{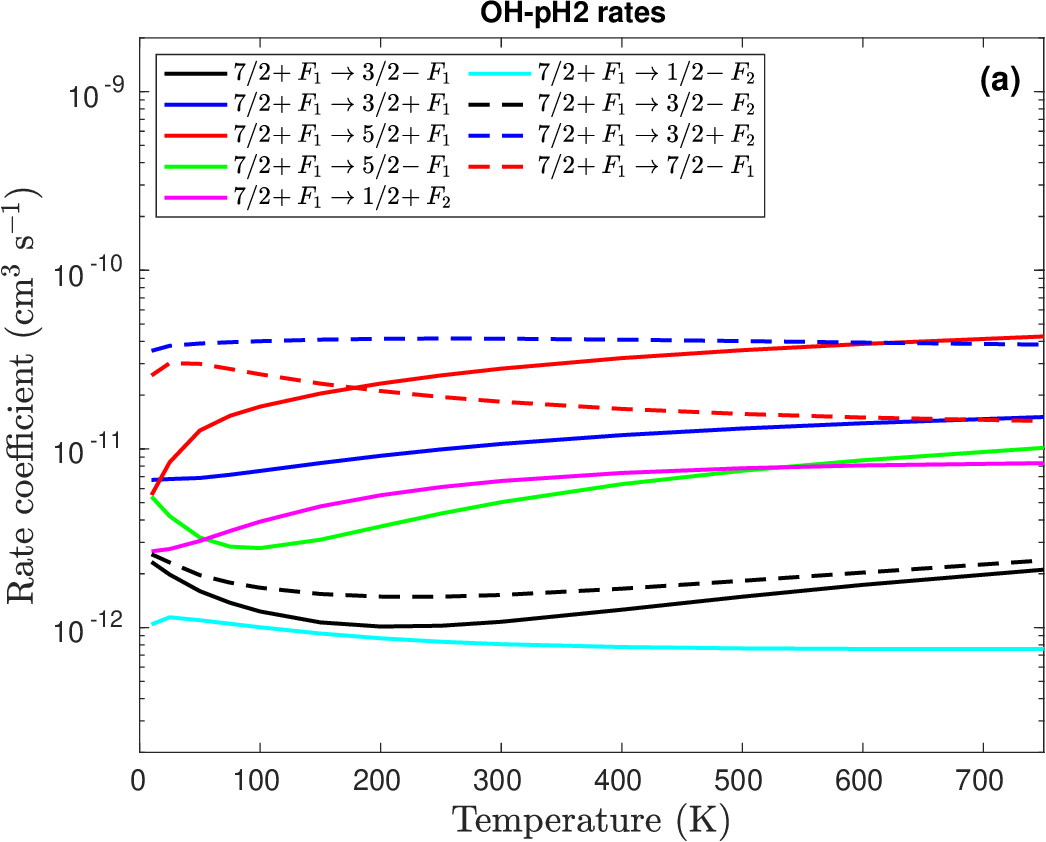} &
\includegraphics*[width=0.48\textwidth]{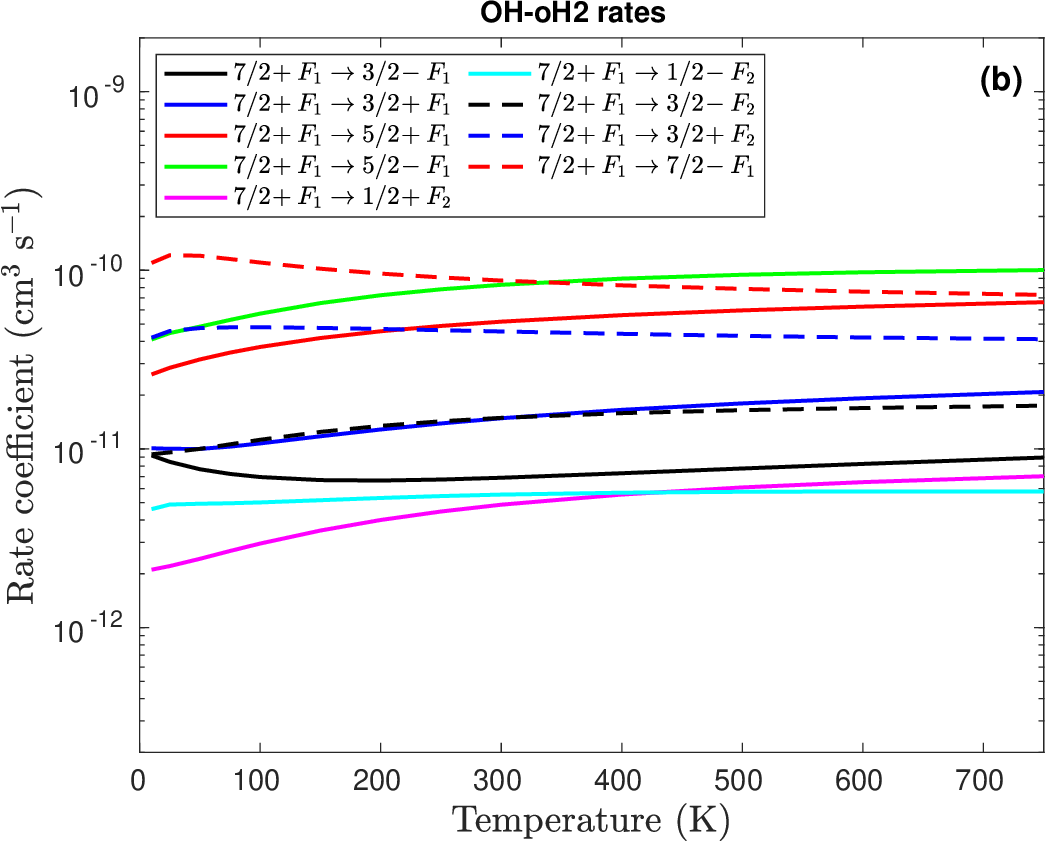} \\
\includegraphics*[width=0.48\textwidth]{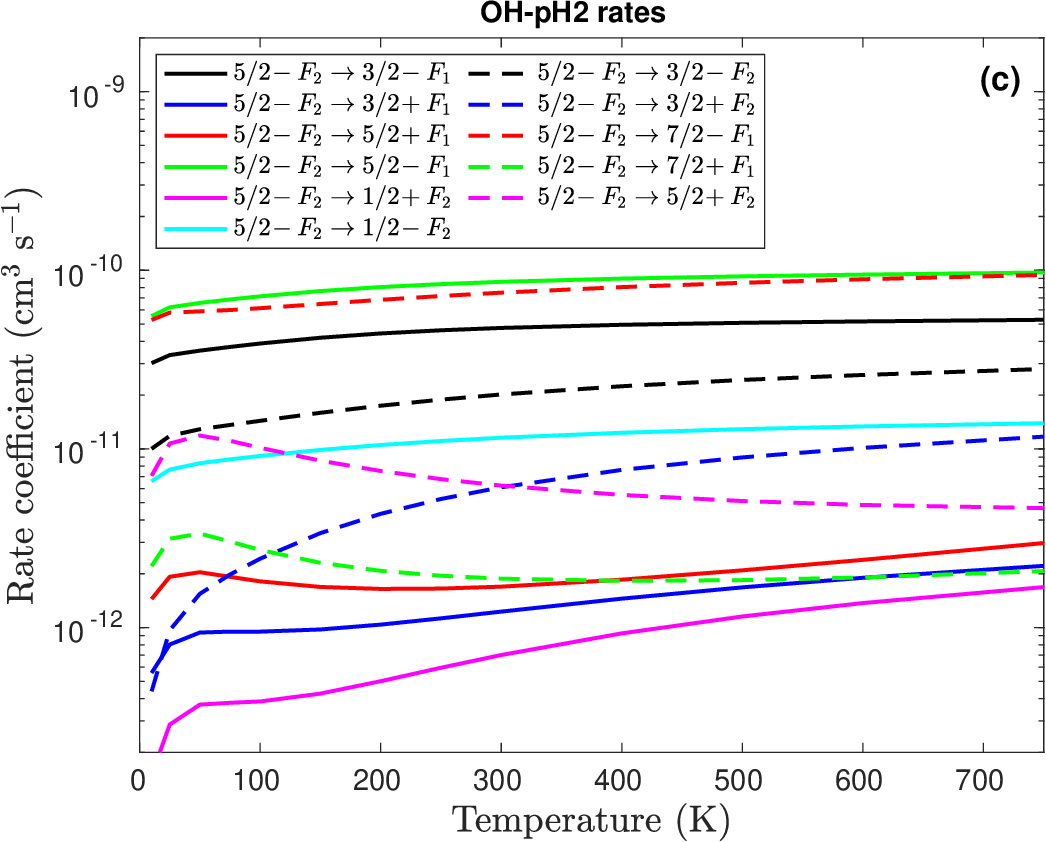} &
\includegraphics*[width=0.48\textwidth]{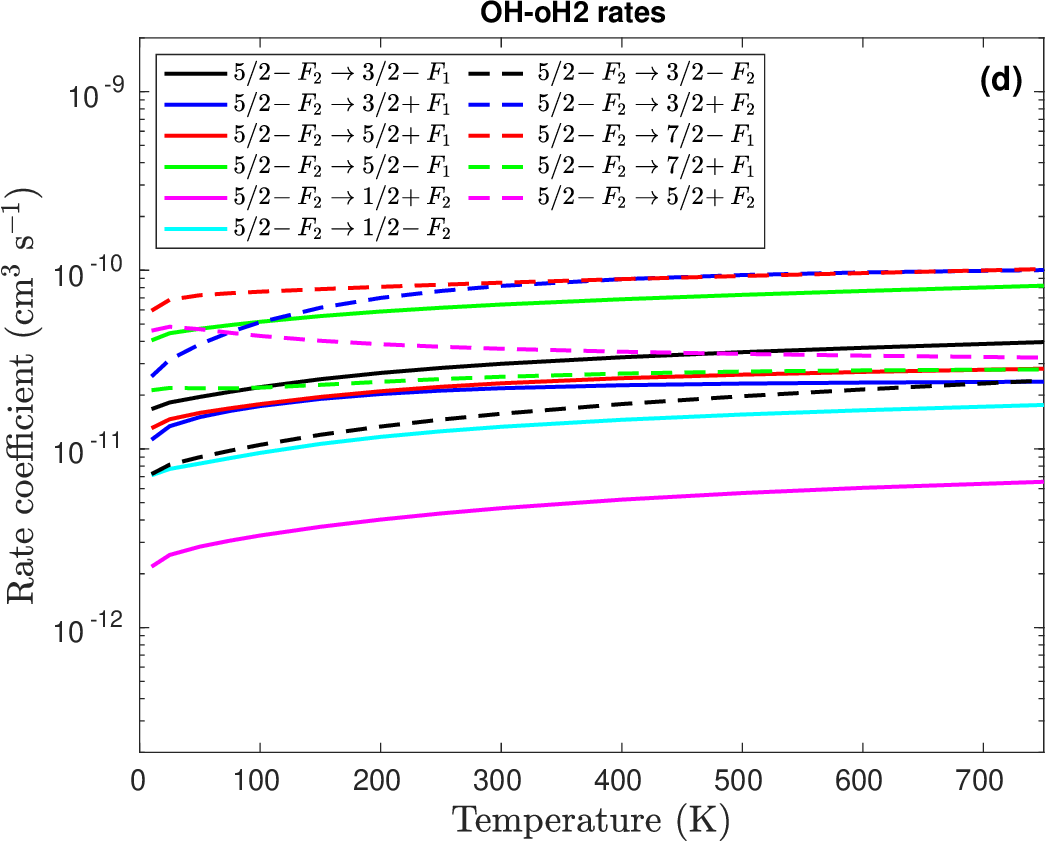} \\
\end{tabular}
\caption{Rate coefficients of OH collisions with para- and ortho-H$_2$
for the same OH transitions of which the cross sections are shown in
Fig.~\ref{fig:cross}. Panels (a), (b), (c), and (d) refer to the
same transitions as in Fig.~\ref{fig:cross}.}
\label{fig:rates}
\end{figure*}

The complete set of calculated state-to-state rate coefficients is made
available through the LAMDA database \citep{Schoeier05,vanderTak20}.
Figures~\ref{fig:cross} and \ref{fig:rates} show some illustrative
results for the cross sections and rate coefficients in collisions of OH
with pH$_2$ and oH$_2$. Panels (a) and (b) in these figures show the
cross sections and rates calculated for an initial state with $j_{\rm
OH} = 7/2$ of parity $p=+1$ (spectroscopic parity $f$) in the lower
spin-orbit manifold $F_1$ for transitions to all lower OH levels, while
panels (c) and (d) show the cross sections and rates calculated for an
initial state with $j_{\rm OH} = 5/2$ of parity $p=-1$ (spectroscopic
parity $f$) in the upper spin-orbit manifold $F_2$, again to all lower
OH levels. The cross sections and rates are summed over all final states
of pH$_2$ and oH$_2$. Figures for higher initial states of OH are shown
and discussed in the Appendix.

Comparisons with the previously calculated cross sections of
\citet{Offer94} are already given in \citet{Schewe15}. Schewe et al.\
concluded that the OH-H$_2$ cross sections which they computed with the
potential of \citet{ma:14} differ significantly from those presented by
\citet{Offer94}, especially for collisions with pH$_2$. Such differences
signify the inaccuracies in the OH-H$_2$ collisional (de-)excitation
rate constants presently in use in various astrophysical applications.
More recent calculations on OH-H$_2$ were published by \ \citet{Klos17},
who used the same potential of \citet{ma:14} as we used in the present
work. They calculated rate coefficients for initial OH states up to
$j=9/2$ and temperatures from 5 to 150\,K, some of which are presented
in graphical form in their paper. The figures we made of our results for
these lower $j$ values and lower temperatures agree precisely with
Fig.~2 in \citep{Klos17}. Also Klos et al.\ concluded that their
OH-H$_2$ rate coefficients are significantly larger than those currently
used for astrophysical modeling.

Our results show that the cross sections and rates of OH transitions in
collisions with oH$_2 (j=1)$ are mostly larger than in collisions
with pH$_2 (j=0)$. This can be explained by the observation that for
pH$_2(j=0)$ the H$_2$ quadrupole moment vanishes by rotational
averaging, so that the important dipole-quadrupole term in the OH-H$_2$
interaction potential is effectively missing. Exceptions are some of
the transitions between the different spin-orbit manifolds $F_1$ and
$F_2$ (also denoted by $\Omega = 3/2$ and $\Omega = 1/2$, respectively).
These transitions are affected by the off-diagonal diabatic OH-H$_2$
potential from which the dipole-multipole terms are missing.

We calculated the cross sections and rates for H$_2$ initially in $j=0$
for pH$_2$ and in $j=1$ for oH$_2$, but at 750\,K initial states of
H$_2$ with higher $j$ are also populated. Therefore, we performed a
series of tests with initial $j_{{\rm H}_2} = 2$ and 3 and found that
the cross sections and rates for both these higher $j$ values of H$_2$
are similar to those in collisions with $j_{{\rm H}_2} = 1$. This is
perhaps surprising since $j_{{\rm H}_2} = 2$ belongs to pH$_2$ while
$j_{{\rm H}_2} = 1$ belongs to oH$_2$, but can be explained by the
observation that the H$_2$ quadrupole moment vanishes only for $j=0$,
but not for any $j>0$. The same conclusion can be drawn from the results
of \citet{Klos17}, who also performed calculations for initial $j_{{\rm
H}_2} = 2$ and 3 and compared the rates with those for initial $j_{{\rm
H}_2} = 0$ and 1 (see Figs.~1 and 2 in their paper). So one can obtain
fairly good estimates of the rate coefficients of OH-H$_2$ collisions
with higher $j_{{\rm H}_2}$ by taking the corresponding values from
OH-H$_2(j=1)$ collisions. Only the rate coefficients  calculated for
initial $j=0$ for pH$_2$ and $j=1$ for oH$_2$ are included in the
LAMDA data base, so for higher $j_{{\rm H}_2}$ the user should take the
oH$_2$ values with $j_{{\rm H}_2}=1$.

In Fig.~\ref{fig:cross} one can observe that the cross sections in
collisions with pH$_2$ as functions of energy show much more structure
than in collisions with oH$_2$. The peaks at energies up to
100~cm$^{-1}$ are caused by scattering resonances, relatively long-lived
quasi-bound states formed during the collision. Furthermore, we find
that at lower energies collisions that are inelastic in H$_2$, i.e., with $\Delta
j_{{\rm H}_2}=\pm 2$ have cross sections that are generally about two
orders of magnitude smaller than collisions with $\Delta j_{{\rm
H}_2}=0$. Figures~\ref{fig:cross}(a) and (c) show, however, that in
collisions with pH$_2$ several cross sections increase steeply at higher
collision energies. This is due to transitions from the initial state
with $j_{{\rm H}_2}=0$ to a final state with $j_{{\rm H}_2}=2$, which
can occur when the OH de-excitation energy plus the collision energy is
sufficient to enable this $\Delta j_{{\rm H}_2}=2$ transition. The
energies where this increase is found are different for different final
OH states because of the different de-excitation energies. These steep
increases do not occur in collisions with oH$_2$ shown in
Figs.~\ref{fig:cross}(b) and (d) since the excitation energy of $j_{{\rm
H}_2}=1$ to 3 is $\simeq 600$~cm$^{-1}$, while the excitation of
$j_{{\rm H}_2}=0$ to 2 requires only $\simeq 360$~cm$^{-1}$.

The rate coefficients displayed in Fig.~\ref{fig:rates} nicely reflect
the effects observed in the cross sections shown in
Fig.~\ref{fig:cross}. The order of the curves corresponding to different
final states is roughly the same in Figs.~\ref{fig:cross} and \ref{fig:rates}, as might be expected
because the rate coefficients were obtained from the corresponding cross
sections with the aid of Eq.~(\ref{eq:rate}). The rate coefficients in
Figs.~\ref{fig:rates}(a) and (c) for collisions with pH$_2$ show
different characteristics. For some of the final OH states the curves
are rather flat as functions of the temperature, but for other final
states the curves clearly rise with the temperature. This rise follows
from the rise of the corresponding cross sections for the latter states
at higher collision energies due to the opening of the $\Delta j_{{\rm
H}_2} = 2$ excitation channel discussed in the preceding paragraph. The
rate coefficients in Figs.~\ref{fig:rates}(b) and (d) for collisions
with oH$_2$ show less pronounced differences as functions of the
temperature, they are mostly quite flat. A few of them rise with the
temperature, but less steeply than the curves for pH$_2$ in
Figs.~\ref{fig:rates}(a) and (c). This corresponds to the trends in the
cross sections in Figs.~\ref{fig:cross}(b) and (d), which mostly
decrease with increasing collision energy but become flatter or start
rising for those final states for which the rate coefficient curves show
rises.

\subsection{Extrapolation to higher initial OH states}
\label{sec:extrap}

\begin{figure}
\begin{tabular}{cc}
\includegraphics*[width=0.46\textwidth]{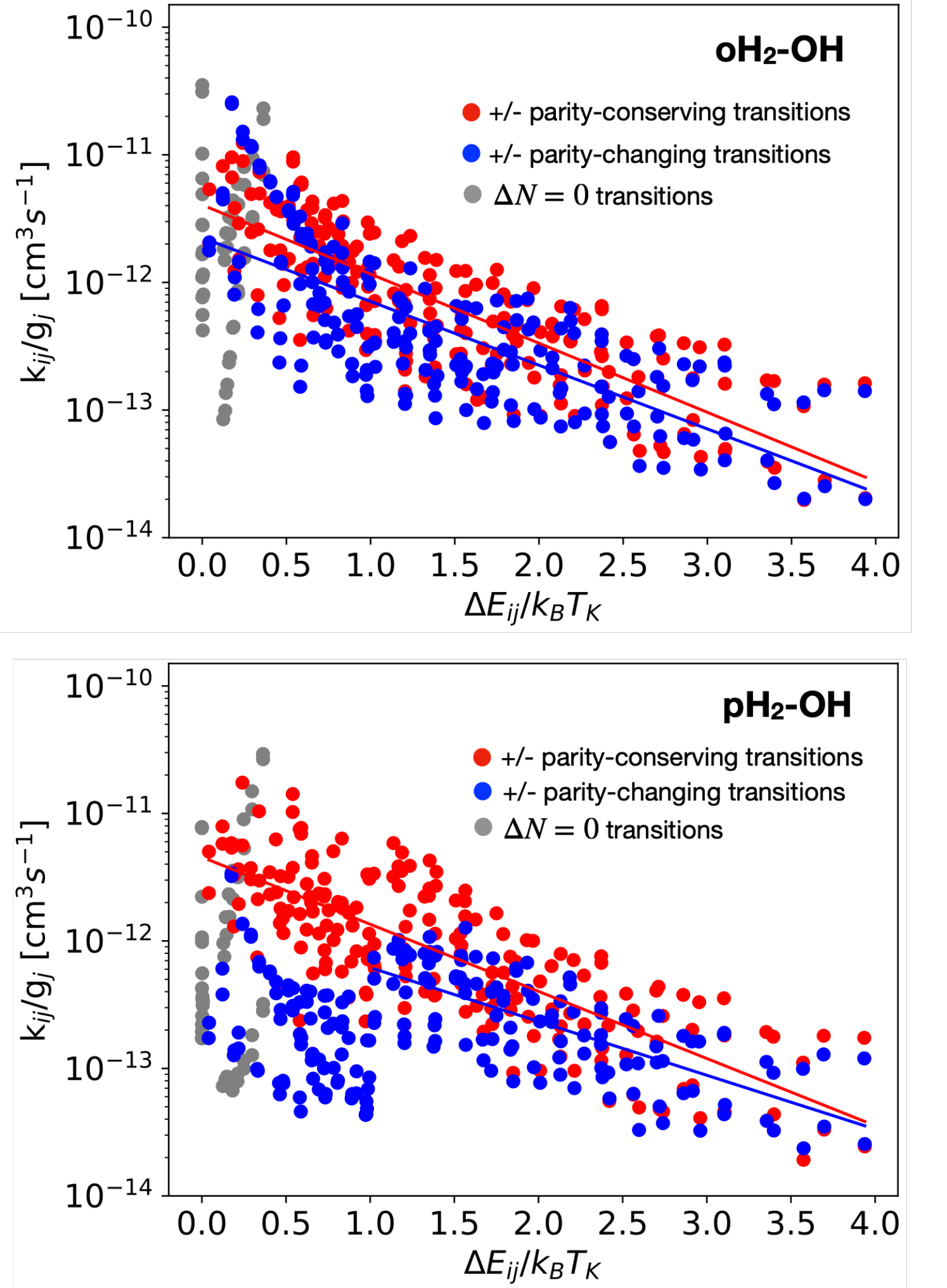}
\end{tabular}
\caption{Collisional rate coefficients normalized by the degeneracy of the lower levels at $T_K=500$~K
as a function of the energy gap $\Delta E_{ij}$ between the upper and
the lower state of the transition.  {Transitions occurring within a single $N$ state are shown as gray dots. Parity-conserving and parity-changing transitions are plotted as red and blue dots, respectively. The straight lines are fits to the parity-conserving (red) and parity-changing (blue) transitions using Eq. (\ref{eq:ProccaLevine}) as an ansatz and excluding $\Delta N=0$ transitions. For parity-changing transitions with pH$_2$, we restricted the fit to $\Delta E_{ij} > 510~$K.}}
\label{fig:rate_coeff_scalings}
\end{figure}

\begin{figure}
\centering
\includegraphics*[width=0.49\textwidth]{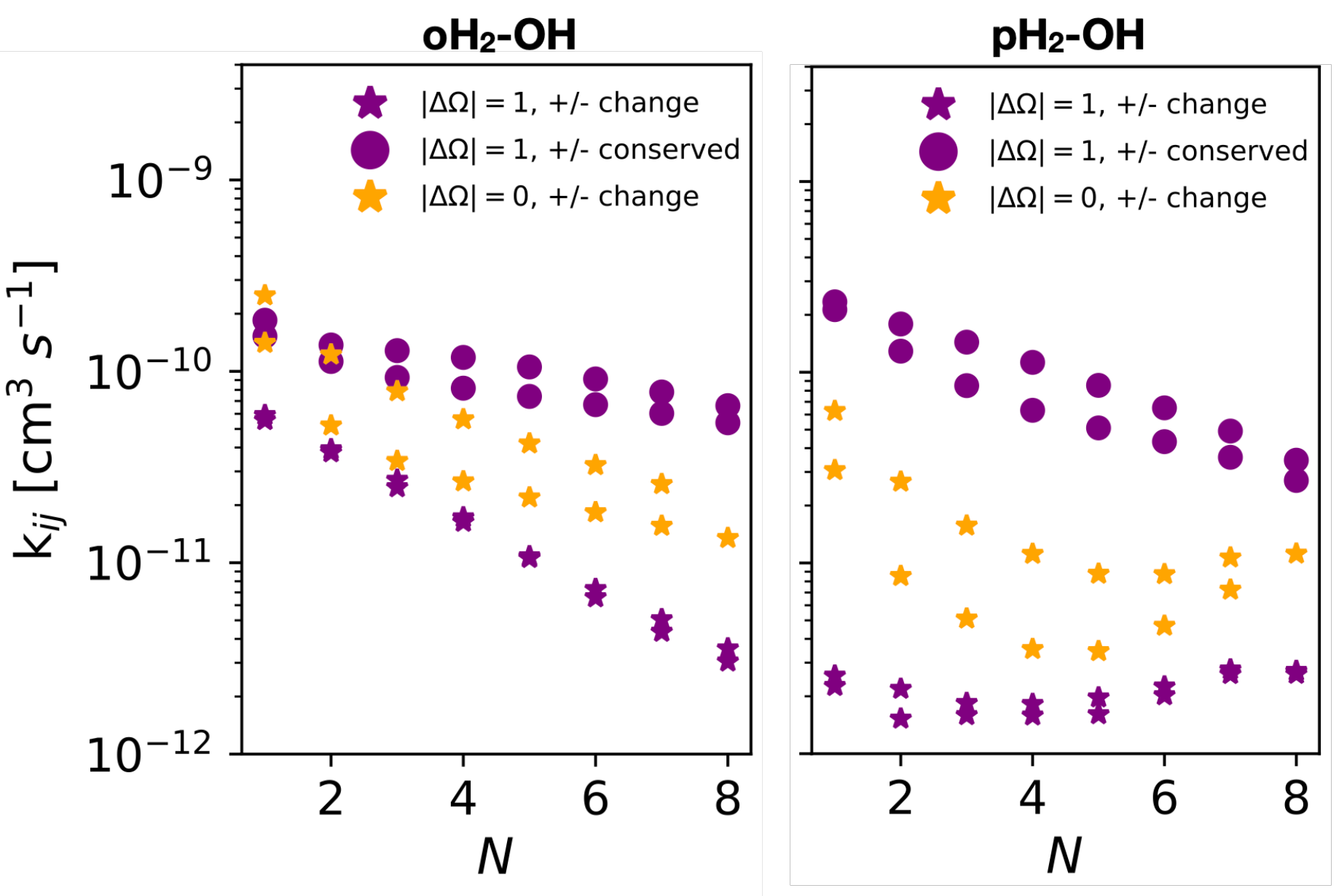}
\caption{Collisional rate coefficients at a kinetic temperature of 500~K
as a function of the $N$ rotational number for the $\Delta N = 0$
collision-induced transitions. These transitions correspond to the lower
energy gap, and their rate coefficients do not follow the correlation
observed in Fig. \ref{fig:rate_coeff_scalings} (see {gray dots}).}
\label{fig:rate_coeff_scalings_DeltaN=0}
\end{figure}

The calculation of the cross sections and rate coefficients for the
initial states of OH up to $j=15/2$ was computationally very demanding.
The brightest rotational transitions starting from these levels
typically emit in the far-IR from $35~\mu$m to $180~\mu$m, the spectral
domain (partly) covered by the \textit{Herschel} Space Observatory. However, the
OH pure rotational lines accessible with JWST MIRI typically probe
rotational levels all the way up to their dissociation threshold
\citep[e.g.,][]{Zannese24}. Inspired by \citet{Faure&Josselin2008}, we
extrapolate the rate coefficient to higher $j$ states from the set of
calculated rate coefficients.

Following the expression proposed by \citet{ProcaLevine1976}, we plot in
Fig.~\ref{fig:rate_coeff_scalings} the downward rate coefficients
normalized by the degeneracy $g_j$ of the lower state as a function of
the energy gap $\Delta E_{ij}$ between the upper and the lower states
$i$ and $j$ involved in the transition. The rate coefficients are well
correlated, confirming the relevance of the \citet{ProcaLevine1976}
expression
\begin{equation}
    k_{ij}(T) = A(T) g_j \exp{ \left( - \Theta_R(T) \Delta E_{ij}/k_B T \right)},
    \label{eq:ProccaLevine}
\end{equation}
where $A(T)$ and $\Theta_R(T)$ are scalar functions of the temperature.
The collisional rate coefficients can then be extrapolated by fitting
the calculated values using Eq.~(\ref{eq:ProccaLevine}) as an ansatz.
However, the rate coefficients exhibit a scatter of $\simeq 1$ dex around the main
trend, reflecting complex dynamical effects accounted for in our
calculations; consequently, a simple extrapolation is expected to
produce typical uncertainties of about one order of magnitude. Large
errors in rate coefficients with low values  are acceptable because they play a negligible
role in the excitation of OH but a good accuracy is required for those with high
values ($k \simeq 10^{-12}$ cm$^{3}$ s$^{-1}$).

The downward rate coefficients do not depend only on the energy gap
$\Delta E_{ij}$ but also on the nature of the collision-induced
transition (see colors in Fig.
\ref{fig:rate_coeff_scalings}). To improve the accuracy of the
extrapolation, we fit separately different subsets of collision-induced
transitions. The $+/-$ parity-conserving transitions are systematically
higher than the parity-changing transitions {(see red versus blue
dots)}. Interestingly,
\citet{Faure&Josselin2008} found that the rate coefficients for H$_2$
collision-induced rotational transitions of H$_2$O are independent of
the rotational quantum numbers when focusing on transitions with
small differences in their rotational number. By contrast, we could not
find robust `high propensity rules' for OH. Transitions with low $\Delta
j$ have higher rate coefficients, but are well in line with the correlation of the
rates with the energy gap. Finally, the collision-induced transitions
occurring within a given $N$ state follow a different pattern with
strong variations within a narrow range of energy gap {(see gray points, Fig. \ref{fig:rate_coeff_scalings})}. In Figure
\ref{fig:rate_coeff_scalings_DeltaN=0} we show that these rate coefficients scale better
with the $N_{\rm up}$ level of the considered transition and that
most rate coefficients vary only weakly for $N>5$.

Based on the observed trends, we extrapolate the rate coefficients by
separating (a) the transitions occurring within a given $N$ state
($\Delta N=0$), (b) the remaining parity-changing transitions, and (c)
the parity-conserving transitions. For the $\Delta N=0$ transitions, we
take the rates computed for the larger $N$ number separating $\Delta
\Omega = \pm 1$ parity-changing, $\Delta \Omega = \pm 1$
parity-conserving, and $\Delta \Omega = 0$ transitions.
For the remaining transitions, we use the ansatz of
Eq.~(\ref{eq:ProccaLevine}). The fits are performed for each temperature
and each state of H$_2$. {To fit the rate coefficients of the parity-changing transitions with pH$_2$, we include only transitions with $\Delta E > 510$~K. This choice avoids rates that are poorly described by our ansatz and that are not relevant for our purposes, since the extrapolated rate coefficients with $\Delta N \neq 0$ typically involve large energy gaps.} The resulting {fits for $T=500~$K are shown as
straight lines} in Fig. \ref{fig:rate_coeff_scalings}. The fits give reasonable
estimates
and we evaluate the error on the extrapolated rates
of high $j$ transitions to be within a factor 5.

\section{Application}
\label{sec:appl}

\begin{figure}
\centering
\begin{tabular}{cc}
\includegraphics*[width=0.45\textwidth]{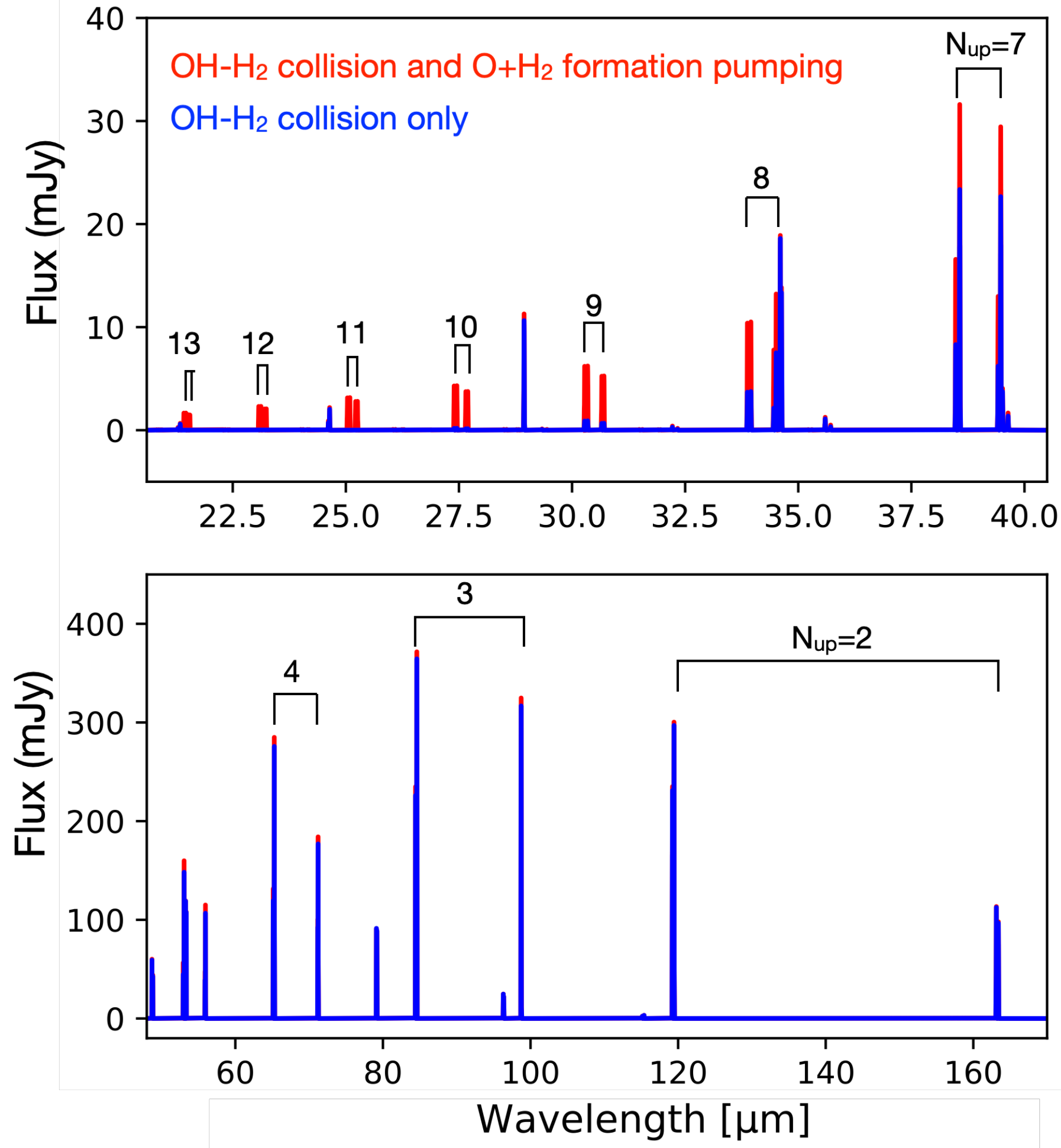}
\end{tabular}
\caption{Synthetic spectra of OH computed with \texttt{GROSBETA} at a spectral resolving power of $\lambda/\Delta \lambda=3000$ assuming an unresolved emission of size $\pi (10~\text{au})^2$ at 140~pc (see text for the other parameters). The red spectrum corresponds to a case where both chemical pumping and collision with H$_2$ are included. The blue spectrum neglects chemical pumping. The similarity of the spectra at long wavelength indicates that collisions with H$_2$ dominate over chemical pumping for $j \lesssim 15/2$. The upper rotational number $N$ is labelled for the pure rotational transition $\Delta N = 1$.}
\label{fig:application}
\end{figure}

OH is abundant under warm ($> 400~$K) and FUV irradiated conditions,
which enhance gas phase formation via O + H$_2$ and promote water
photodissociation, thereby preventing oxygen from being locked into
water. Under these conditions, the rotational levels are expected to be
excited not only by collisions but also by formation pumping. Focusing
on $N<20$ lines, the most compelling formation pumping is via O+H$_2$
\citep{Liu2000,Carr14}. Today, the only evidence of excitation by
chemical pumping comes from the vibrational emission \citep{Zannese24},
even though non-thermal distributions of rotationally excited OH support the relevance of this process in the mid-IR emission of OH
\citep{Carr14}.

Here, we study how collisions with H$_2$ can compete with chemical
pumping using the single-zone model \texttt{GROSBETA} developed by J. H.
Black \citep{Tabone21}. This code is an advanced version of the widely
used \texttt{RADEX} code \citep{vanderTak07} with the option of
including the effect of formation pumping. Following the formalism of formation pumping described in
\citet{Tabone21} and \citet{Zannese24}, the model is controlled by four
parameters: the column density of OH $N({\rm OH})$ [cm$^{-2}$], the
column density of OH formed per unit of time $R_f$ [cm$^{-2}$~s$^{-1}$],
the total hydrogen number density $n_{\rm H}$ [cm$^{-3}$], and the kinetic
temperature $T_K$. To model warm molecular gas typically found in
interstellar PDRs \citep{Parikka17}, dense interstellar shocks
\citep{Karska2013}, or at the surface of protoplanetary disks
\citep{Fedele13}, we chose a temperature of $T_K = 1000$\,K, a column
density of $N({\rm OH}) = 10^{15}~$cm$^{-2}$, and a total hydrogen
density of $n_{\rm H} = 10^8$cm$^{-3}$. We further assume that the H$_2$
levels are populated according to a Boltzmann distribution at the gas
temperature. Following Sec.~\ref{sec:results}, the collisional rate
coefficients for $j_{{\rm H}_2}>1$ are assumed to be those for $j_{{\rm
H}_2}=1$. The collisional rate coefficients are not extrapolated in
temperature, and we adopted those at $T_K=750$\,K.

The formation rate of OH can be written as $R_f=
x({\rm O}) x({\rm H_{\text{2}}}) /x({\rm {\rm OH}}) \, n_{\rm H} \, k_{\rm chem} \, N({\rm OH})$,
where $ k_{\rm chem}$ is the OH formation rate coefficient and $x$
designates the abundance of the species with respect to the total number
of hydrogen atoms. Thermochemical models show that the O/OH abundance
ratio is at least 20 in regions where OH is abundant
\citep{Glassgold09,Zannese2023}. In addition, the chemical formation
rate at $1000~$K is $k_{\rm chem} \simeq 2 \times 10^{-13}$cm$^{3}$ s$^{-1}$. {Therefore, we choose a value of $R_f = 1 \times 10^{12}$~cm$^{-2}$ s$^{-1}$ which corresponds to $x({\rm O})/x({\rm {\rm OH})} = 100$ for the adopted values of $T_K$ and $ N({\rm OH})$ and assuming an H$_2$ dominated gas.} {The distribution of nascent OH following O+H$_2$ is a Boltzmann distribution at $T=2,000~$K, in line with the distribution of the rotational levels computed by \citet{Zannese24} at a $T_K=1,000$~K \citep[Zannese, private communication from][]{Veselinova2021}.}

Figure \ref{fig:application} shows the importance of the calculated rate
coefficients on the modeling of OH emission. Comparing the model with a
model where chemical pumping is switched off, it is apparent that
collisional excitation dominates the population of OH up to rotational
number of about $j \simeq 15/2$ (bottom panel). Shortward of {35~$\mu$m},
the OH lines are mostly controlled by chemical pumping. These lines
correspond to pure rotational lines with $j \gtrsim 15/2$, except the
bright lines, which are cross-ladder transitions starting from low $j$
states. In other words, the excitation of OH is dominated by chemical
pumping for high $j$ and collisions with H$_2$ for lower $j$. This
distinction is because chemical pumping directly produces OH in
rotationally excited states, whereas it requires multiple collisional
transitions to populate high $j$ lines.

The model presented here is only illustrative. A more realistic model
requires adopting a nascent distribution of OH, which could deviate from
a Boltzmann distribution due to the complex dynamics of the O+H$_2$
reaction \citep{Veselinova2021}. Also, collisions with atomic hydrogen
and He have been omitted for simplicity. Lastly, OH prompt emission
produced by water photodissociation is ignored. This process populates
much more excited levels of OH that are not affected by collisions, but
they can impact the intermediate $j$ states following a radiative
cascade.

\section{Conclusions}
\label{sec:concl}
OH is a cornerstone molecule in the chemistry of interstellar and
circumstellar media and is ubiquitously detected in spectra from radio
to near infrared wavelengths. Its excitation is the result of several
microphysical processes that can serve as a unique probe of the physical
and chemical conditions in space. In this work, we expanded the set of
collisional rate coefficients to include rotational levels up to
$j=15/2$ and up to a temperature of 750~K, typical of the conditions
where OH is abundant. These rate coefficients are obtained from
well-converged close-coupling quantum scattering calculations and on the
basis of our experience with such calculations and with \textit{ab
initio} calculated intermolecular potentials
\citep{jongh:20,kuijpers:25}, they are estimated to be accurate to
better than a few percent. They are further extrapolated to higher
rotational levels of OH based on scaling relations inferred from the
calculated dataset. The data, provided to the community via the LAMDA
database, will open up many applications. As a proof of concept, we
present \texttt{GROSBETA} models with a simplified implementation of
chemical pumping of OH and demonstrate that collisions with H$_2$ are
fundamental for analyzing OH excitation and therefore obtain direct
access to the warm gas in interstellar and circumstellar media.

\begin{acknowledgements}
The authors are grateful to Robert Forrey for his careful review of the manuscript. BT thanks A.~Faure for valuable discussions about the extrapolation of
collisional rate coefficients and J. H. Black for sharing the \texttt{GROSBETA} code. This work was supported by the Programme
National PCMI of CNRS/INSU with INC/INP cofunded by CEA and CNES.
Astrochemistry in Leiden is supported by funding from the European
Research Council (ERC) under the European Union's Horizon 2020 research
and innovation program (grant agreement No. 291141 MOLDISK), and by the
Danish National Research Foundation through the Center of Excellence
``InterCat'' (DNRF150).

\end{acknowledgements}
  \bibliographystyle{aa}

\bibliography{biblio_ohh2.bib}

\begin{appendix}
\section{Rate coefficients for higher initial states of OH}
\begin{figure*}
\begin{tabular}{cc}
\includegraphics*[width=0.48\textwidth]{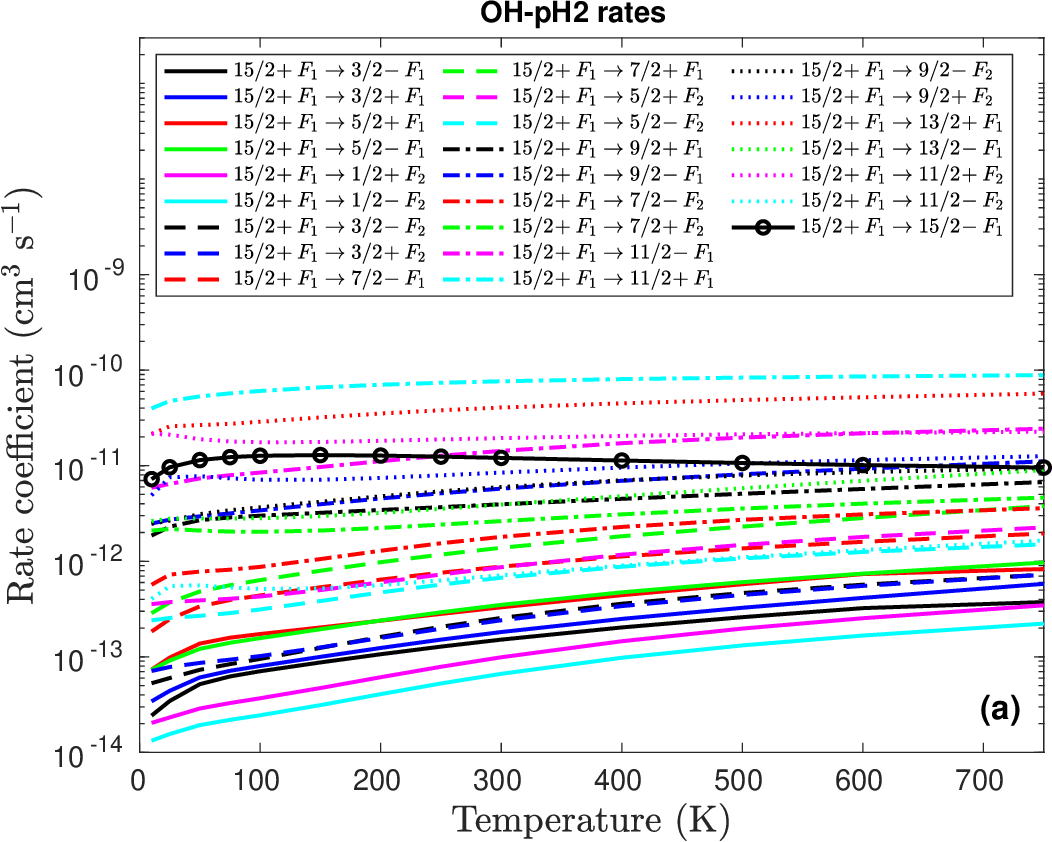} &
\includegraphics*[width=0.48\textwidth]{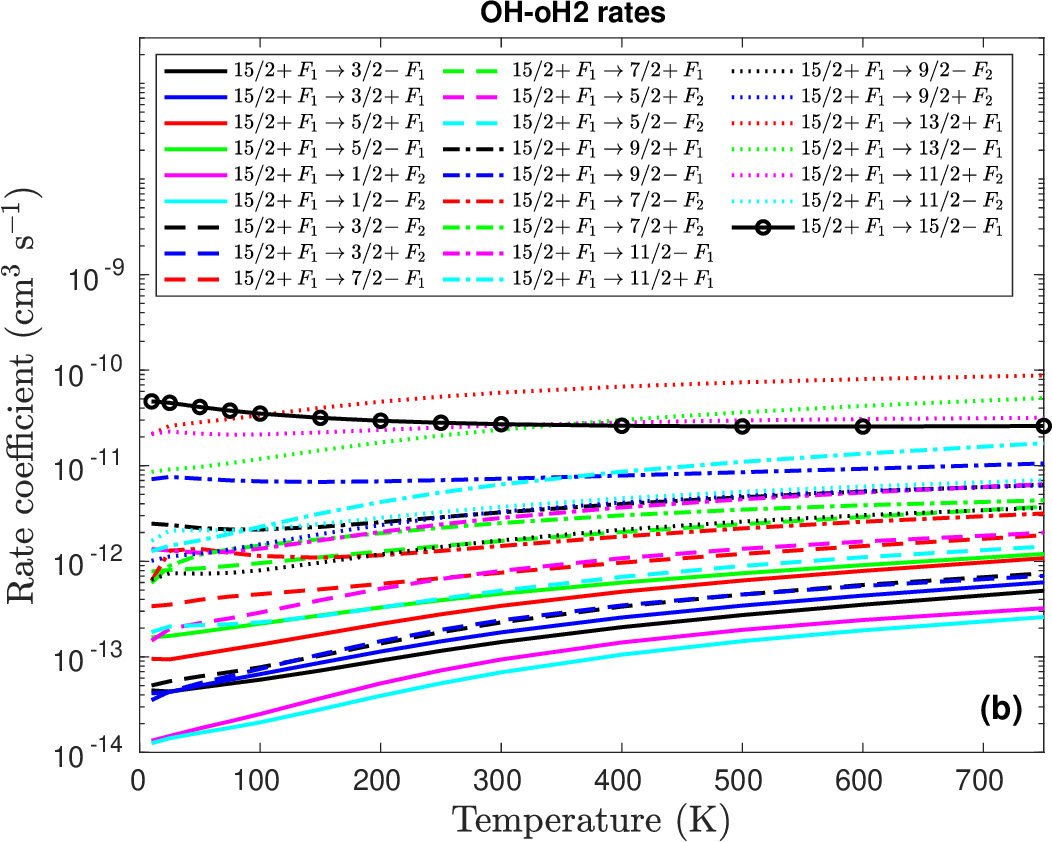} \\
\includegraphics*[width=0.48\textwidth]{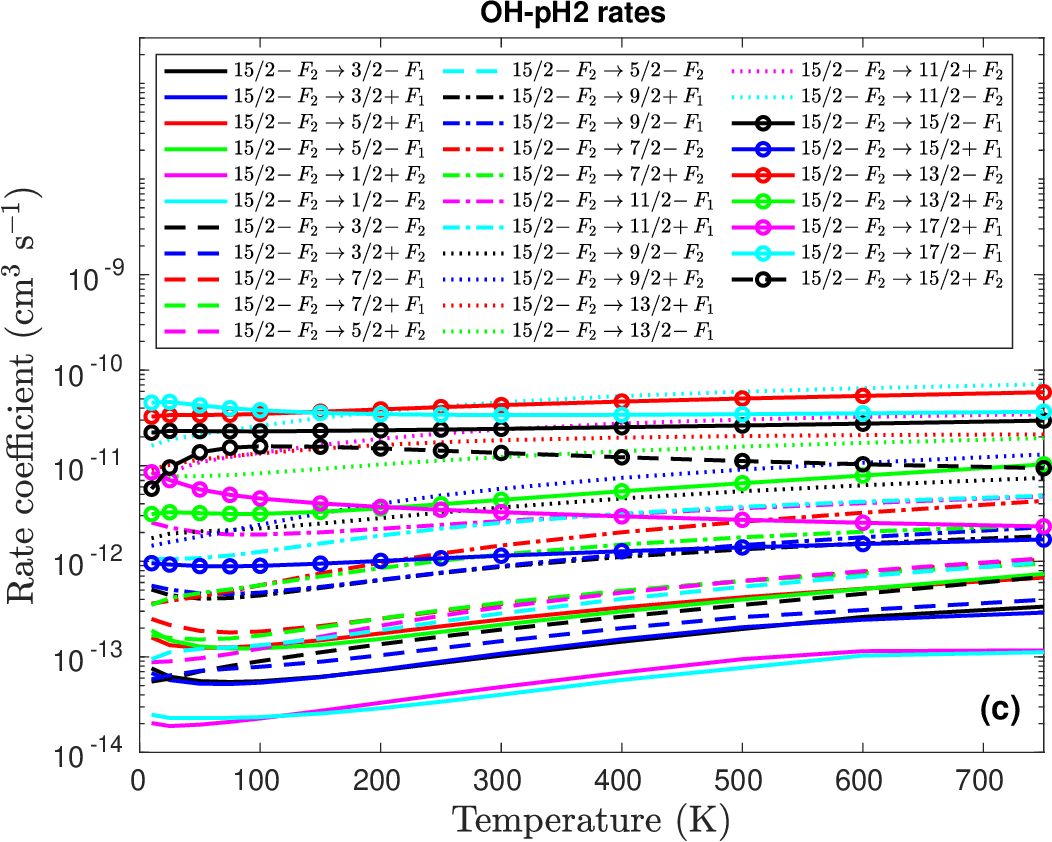} &
\includegraphics*[width=0.48\textwidth]{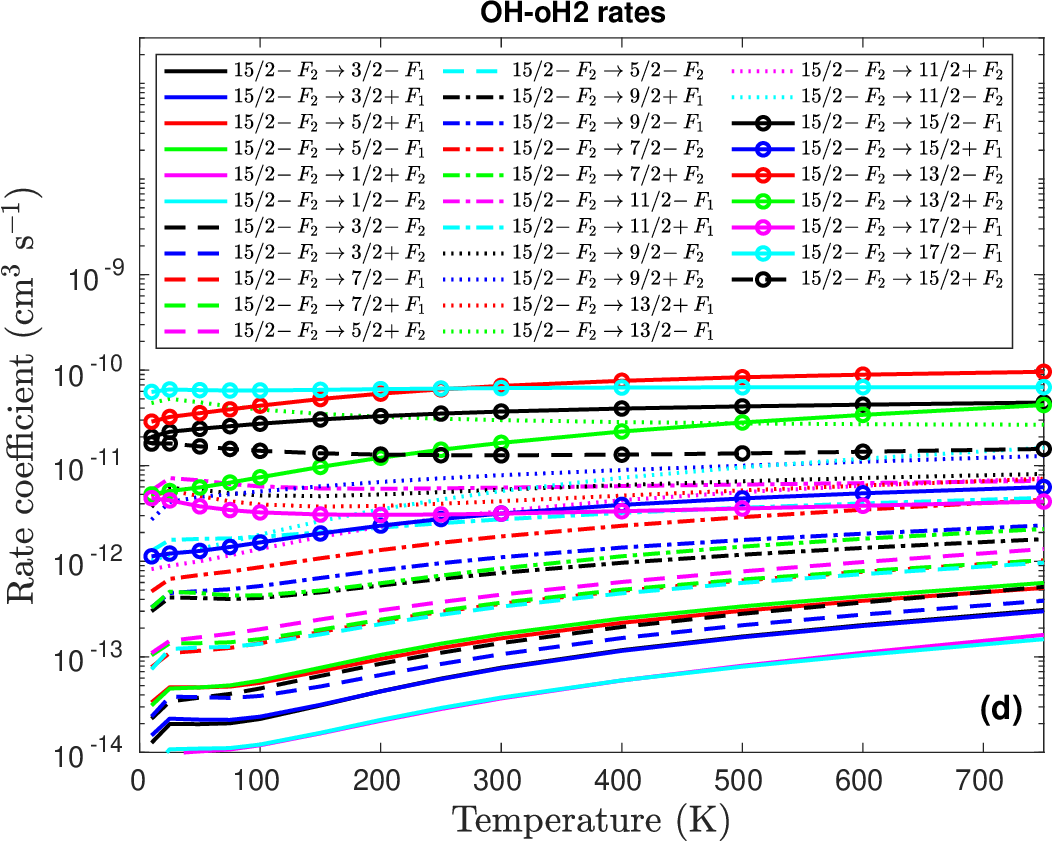}
\end{tabular}
\caption{Rate coefficients of OH transitions from the $F_1$ state with
$j=15/2$ and parity $p=+1$ (spectroscopic parity $f$) [panels (a) and
(b)] and the $F_2$ state with $j=15/2$ and parity $p=-1$ (spectroscopic
parity $e$) [panels (c) and (d)] to all lower levels of OH and summed
over all final H$_2$ states. Panels (a) and (c) refer to collisions with
pH$_2$, panels (b) and (d) to collisions with oH$_2$.}
\label{fig:appen}
\end{figure*}
In order to see trends in the rate coefficients with increasing
rotational states of OH Fig.~\ref{fig:appen} shows the results for
initial $j_{\rm OH}=15/2$, which is the highest rotational state of OH for
which we directly calculated the rates with the quantum scattering
approach. We observe that the largest rates for these higher initial
states of OH have more or less the same magnitude as the largest rate
coefficients for the lower initial $j_{\rm OH}$ values in
Fig.~\ref{fig:rates}. The smallest rate coefficients in
Fig.~\ref{fig:appen} are much smaller than those in
Fig.~\ref{fig:rates}, however, because the energy gaps between
these high initial OH states and the lowest final states are much larger
than in the case of lower initial states. The temperature dependence
of the rate coefficients does not show striking differences.
\end{appendix}
\end{document}